\documentclass[usenatbib]{mnras}

\usepackage{graphicx}
\usepackage{amsmath}
\usepackage{amssymb}
\usepackage{multicol}
\usepackage{color}
\usepackage{blkarray}

\newcommand{\be}{\begin{equation}}
\newcommand{\ee}{\end{equation}}
\newcommand{\bea}{\begin{equation}\begin{aligned}}
\newcommand{\eea}{\end{aligned}\end{equation}}
\newcommand{\td}{{\rm d}}

\makeatletter
  \def\title@font{\Large\bfseries}
  \let\ltx@maketitle\@maketitle
  \def\@maketitle{\bgroup%
    \let\ltx@title\@title%
    \def\@title{\resizebox{\textwidth}{!}{%
      \mbox{\title@font\ltx@title}%
    }}%
    \ltx@maketitle%
  \egroup}
\makeatother

\title[Lensing of GWs as a probe of compact DM]{Lensing of gravitational waves as a probe of compact dark matter}

\author[Urrutia and Vaskonen]{
Juan Urrutia\thanks{juan.urrutia@e-campus.uab.cat} and
Ville Vaskonen\thanks{vvaskonen@ifae.es}
\\
Institut de Fisica d'Altes Energies, The Barcelona Institute of Science and Technology, Campus UAB, 08193 Bellaterra, Barcelona
}

\pubyear{2021}

\begin{document}
\label{firstpage}
\pagerange{\pageref{firstpage}--\pageref{lastpage}}
\maketitle

\begin{abstract}
We study gravitational lensing of gravitational waves from compact object binaries as a probe of compact dark matter (DM) objects such as primordial black holes. Assuming a point mass lens, we perform parameter estimation of lensed gravitational wave signals from compact object binaries to determine the detectability of the lens with ground based laser interferometers. Then, considering binary populations that LIGO-Virgo has been probing, we derive a constraint on the abundance of compact DM from non-observation of lensed events. We find that the LIGO-Virgo observations imply that compact objects heavier than $M_l = 200M_\odot$ can not constitute all DM and less than $40\%$ of DM can be in compact objects heavier than $M_l = 400M_\odot$. We also show that the DM fraction in compact objects can be probed by LIGO in its final sensitivity for $M_l > 40M_\odot$ reaching $2\%$ of the DM abundance at $M_l > 200M_\odot$, and by ET for $M_l > 1M_\odot$ reaching DM fraction as low as $7\times 10^{-5}$ at $M_l > 40M_\odot$. 
\end{abstract}

\begin{keywords}
gravitational waves -- black hole mergers -- neutron star mergers -- dark matter -- gravitational lensing
\end{keywords}

\section{Introduction}

Gravitational wave (GW) observations have proven to provide very powerful probes of  new physics. For example, the constraints already obtained on the speed of GWs~\citep{LIGOScientific:2017zic}, the graviton mass~\citep{LIGOScientific:2019fpa} and the nuclear matter equation of state~\citep{Annala:2017llu} demonstrate the significance of the LIGO-Virgo observations~\citep{LIGOScientific:2018mvr,LIGOScientific:2020ibl}. In near future, we expect even more sensitive GW detectors, such as ET~\citep{Punturo:2010zz,Reitze:2019iox}, CE~\citep{Reitze:2019iox}, LISA~\citep{LISA:2017pwj}, AION-MAGIS~\citep{Graham:2017pmn,Badurina:2019hst} and AEDGE~\citep{AEDGE:2019nxb,Badurina:2021rgt}, to guide us towards solving various open question in modern physics.

One such question is about the origin of dark matter (DM). As the theoretically favoured particle DM candidates have become increasingly tightly constrained by experiments (see e.g.~\cite{Arcadi:2017kky}), other possibilities, such as primordial black holes (PBHs), warrant further studies. Already from the LIGO-Virgo observations, strong constraints have been derived on the abundance of $\mathcal{O}(1M_\odot) - \mathcal{O}(100M_\odot)$ PBHs based on their merger rate~\citep{Sasaki:2016jop,Raidal:2017mfl,Ali-Haimoud:2017rtz,Raidal:2018bbj,Vaskonen:2019jpv,DeLuca:2020qqa,Hutsi:2020sol}. However, it still remains possible that some of the LIGO-Virgo events are of primordial origin~\citep{Hall:2020daa,Hutsi:2020sol,DeLuca:2021wjr,Franciolini:2021tla}. Moreover, it has been shown that future more sensitive experiments provide very powerful probes of PBH binaries in the mass range from $10^{-5}M_\odot$ to $10^8M_\odot$~(see e.g.~\cite{DeLuca:2021hde,Pujolas:2021yaw,Mukherjee:2021ags}), and formation of PBHs in particular at masses below $10^{-5}M_\odot$~(see e.g.~\cite{Inomata:2018epa,Wang:2019kaf,Romero-Rodriguez:2021aws,Badurina:2021rgt}).

The abundance of PBHs, and compact DM objects in general, can be probed through gravitational lensing. Searches of microlensing events indicate that over a broad mass range from $\mathcal{O}(10^{-10}M_\odot)$ to $\mathcal{O}(1M_\odot)$ compact objects constitute less than $1 - 10\%$ of DM~\citep{Griest:2013aaa,Macho:2000nvd,EROS-2:2006ryy,Niikura:2017zjd,Niikura:2019kqi}. These searches are done by observing a large number of stars and looking for amplifications in their brightness that would be caused by massive bodies passing through the line of sight. The duration of the lens event is characterized by the Einstein radius and velocity of the lens. For masses higher than $\mathcal{O}(1M_\odot)$ the long duration of the possible lens events prevents the microlensing searches from probing the abundance of compact DM objects. Instead, for compact objects lighter than $\mathcal{O}(10^{-10}M_\odot)$ their Schwarzschild radii is smaller than the wavelengths at which the searches are done, and the wave optics effects make the possible lensing events less pronounced.

In the same way as electromagnetic waves, also GWs get lensed by compact objects. There is, however, two major differences when looking at lensing of GWs from compact object binaries instead of lensing of stellar light: First, the GW signal from a compact object binary is coherent and therefore the lensing effect can be searched from the signal even if the lensing event would last longer than the observation time. Second, the characteristic wavelength of the GW signals is very long and the wave optics effects become relevant for much heavier lenses than in the case of the microlensing searches. Therefore, searches of lensed GW signals complement the searches of stellar microlensing events, providing a probe of heavier lens objects.

In this paper we study the possibility of probing compact DM objects through gravitational lensing of GW signals from black hole (BH) and neutron star (NS) binaries. We focus on the prospects of the ground based GW laser interferometers LIGO-Virgo in their current sensitivity (O3), LIGO in its final sensitivity and ET. These experiments are sensitive to frequencies around $\mathcal{O}(1 - 1000\,{\rm Hz})$ and give the best potential for probing PBHs down to $\mathcal{O}(1 - 100M_\odot)$. Other forthcoming experiments, such as LISA, are sensitive to lower frequencies and, because of the wave optics effects, they can probe only heavier lenses than the ground based laser interferometers.

Similar studies have been performed by \cite{Jung:2017flg}, \cite{Liao:2020hnx} and \cite{Wang:2021lij}. The most important difference between our analysis and the analyses in these references is in how the detectability of the lens is determined. Whereas \cite{Jung:2017flg} and \cite{Liao:2020hnx} use a simple least-squares fit method and \cite{Wang:2021lij} use fitting factor and odds ratio analyses, we determine the detectability of the lens by requiring that the lens parameters can be measured from the waveform. We do this using Fisher analysis, through which we can estimate the accuracy at which different parameters of the waveform, including the lens parameters, can be measured. We find that geometrical optics can be used to estimate the expected number of detectable lensed GW events. We calculate of the expected number of detectable lensed GW events by integrating over the merger rate of the source population accounting for the detectability of the signal, for the probability for the event to be lensed, and for the detectability of the lens object. As the sources we consider the binary BH and NS populations that the LIGO-Virgo detectors are probing, modeled with a rate that follows the star formation rate with a power-law delay time distribution, and a truncated power-law mass distribution in the case BHs and a monochromatic mass distribution in the case of NSs. Finally, we study how small abundances and masses of the compact DM objects can be probed with the current sensitivity of LIGO-Virgo, and the design sensitivities of LIGO and ET.

Our results agree qualitatively with those found in Refs. ~\citep{Jung:2017flg,Liao:2020hnx,Wang:2021lij}. However, our improved analysis of the detectability of the lens objects puts the results on a stronger footing, and our results differ in two important ways from those obtained in the above references: First, and most importantly, we find that the LIGO-Virgo observations already put constraints on the abundance of compact lens objects heavier than $200M_\odot$. Second, we find a similar reach for the design sensitivities of LIGO and ET than what was found in the earlier literature: We find that LIGO and ET can probe the abundance of compact objects heavier than $40M_\odot$ and $1M_\odot$, respectively.

\section{Lensing of gravitational waves}

We start by briefly reviewing the calculation of lensing of GWs presented in~\citep{Takahashi:2003ix}. To describe lensing of GWs we consider a lens potential $U$ in FLRW background, 
\be \label{eq:lineelement}
    \td s^2 = -\left(1+2U(\vec{r})\right) \,\td t^2 + a(t)^2 \left(1-2U(\vec{r})\right) \,\td \vec{r}^{\,2} \,,
\ee
where $a(t)$ is the scale factor, and $t$ and $\vec{r}$ are the time and spatial coordinates. A GW is described by a small perturbation $h_{\mu\nu}$ over the background metric $g_{\mu\nu}$ described by Eq.~\eqref{eq:lineelement}. The amplitude $h$ of the perturbation follows the equation of motion
\be
    \partial_\mu \left(\sqrt{-g} g^{\mu\nu} \partial_\nu h \right) = 0 \,.
\ee
Lensing can be approximated as a local event occurring near the lens object. Therefore we can neglect the time dependence of the scale factor and in Fourier space the equation of motion for $h$ can be written as
\be \label{Fourier}
    \left(a_l^{-2}\nabla^2 + \omega^2\right) \tilde h = 4 \omega^2 U \tilde h \,,
\ee
where $\omega = 2\pi f$, $\tilde h$ is the Fourier transform of $h$, and $a_l$ denotes the scale factor at the distance of the lens. 

\begin{figure*}
    \centering
    \includegraphics[scale=0.13]{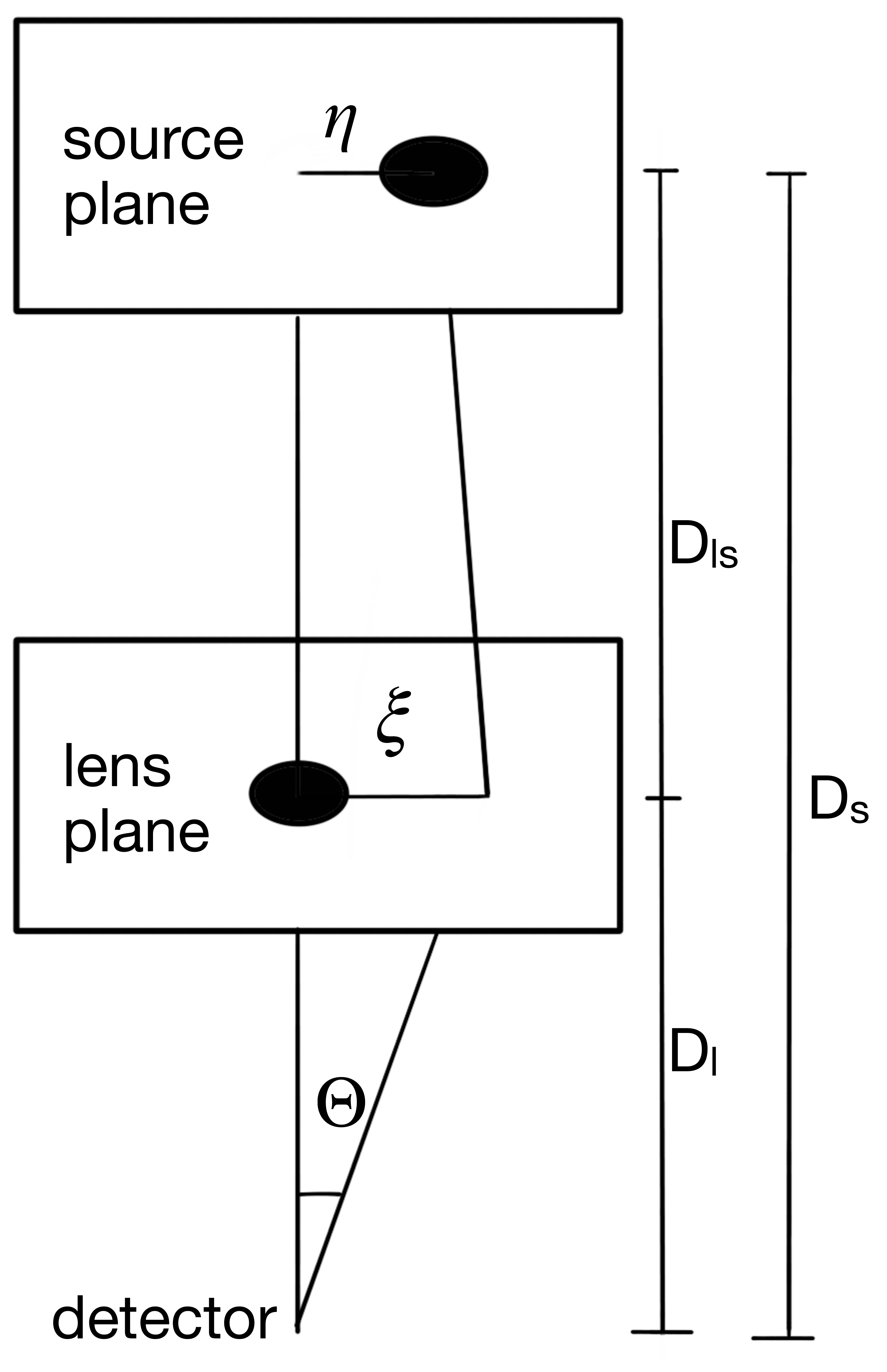} 
    \hspace{1.2cm}  
    \includegraphics[width=0.52\textwidth]{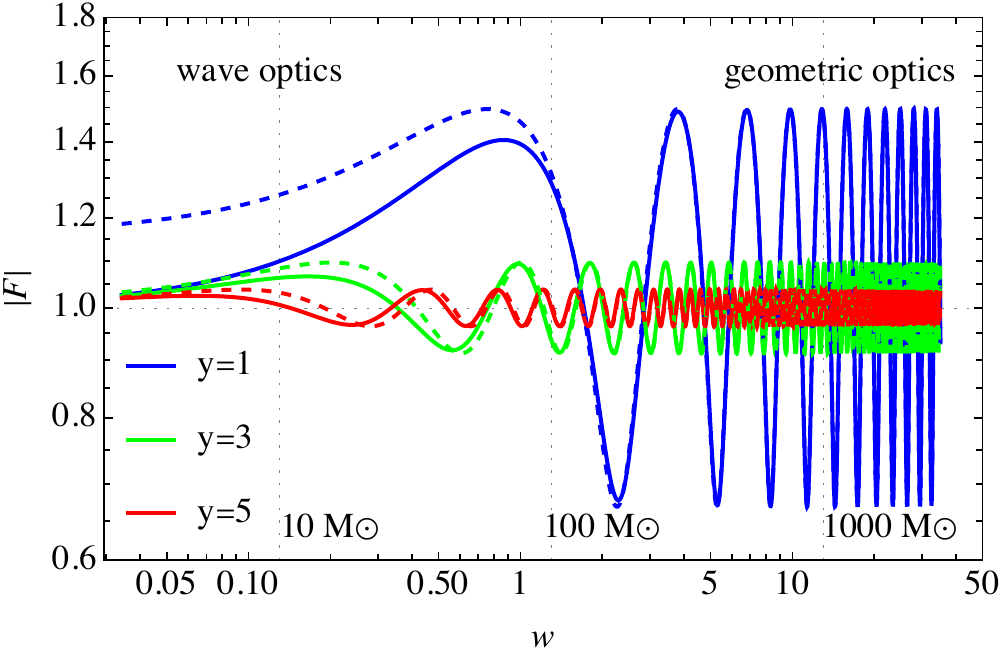}
    \caption{\emph{Left panel:} A schematic picture of the source-lens-detector system. \emph{Right panel:} Lensing amplification factor $|F|$ for a point mass lens as a function of the dimensionless frequency $w$. The vertical dotted lines show the cut-off values of $w$ for a binary of $20M_{\rm \odot}$ BHs at a luminosity distance $D_L=1.5\rm Gpc$ considering different values of the lens mass $M_{lz}$ indicated in the plot.}
    \label{fig:amplification}
\end{figure*}

The equation of motion~\eqref{Fourier} can be re-expressed in terms of the amplification factor
\be \label{amplificationfactor}
    F(f) \equiv \frac{\tilde{h}(f)}{\tilde{h}_0(f)} \,,
\ee
where $\tilde h_0$ corresponds to unlensed ($U=0$) signal. Its solution in the thin-lens approximation is given by\footnote{We assume a point like source. For the GW signals from compact object binaries in the frequency and mass ranges considered in this work, the finite source size can be neglected~\citep{Matsunaga:2006uc}.}
\be \label{Fexpand}
    F(f) = \frac{D_s \xi_0^2 (1+z_l) f}{i D_l D_{ls}} \int d^2x e^{i 2\pi f t_d(\vec{x},\vec{y})} \,,
\ee
where, as illustrated in the left panel of Fig.~\ref{fig:amplification}, $D_s$, $D_l$, and $D_{ls} = D_s - (1+z_l)/(1+z_s) D_l$ are the angular diameter distances to the source, to the lens, and between the lens and the source, $z_l$ and $z_s$ are the redshift of the lens and the source, $\xi_0$ is a normalization constant, $t_d$ is the arrival time delay of the wave to the observer, and
\be
    \vec{x} = \frac{\vec\xi}{\xi_0}\,, \qquad 
    \vec{y} = \frac{D_l}{D_s}\frac{\vec\eta}{\xi_0}\,.
\ee
The vector $\vec\eta$ gives the source position in the source plane and the vector $\vec\xi$ the impact parameter in the lens plane. The arrival time delay is
\be
    t_d(\vec{x},\vec{y}) = \frac{D_s \xi_0^2 (1+z_l)}{D_l D_{ls}} \left[ \frac12\left|\vec{x}-\vec{y}\right| - \psi(\vec{x}) + \phi_m(\vec{y}) \right] \,,
\ee
where the function $\psi(\vec{x})$ is the deflection potential and $\phi_m(\vec{y})$ is chosen such that the minimum of $t_d(\vec{x},\vec{y})$ is zero. 

For a point mass lens the characteristic scale is given by the Einstein ring radius, and $\xi_0 = D_l \theta_E$, where $\theta_E$ is the Einstein angle,
\be \label{Einsteinradius}
    \theta_{E}^2 = 4M_l \frac{D_{ls}}{D_lD_s} \,.
\ee
The deflection potential of a point mass lens is $\psi(\vec{x}) = \ln x$, for which Eq.~\eqref{Fexpand} can be analytically integrated to get
\bea \label{pointamplitud}
    F(f) = &\exp\left[\frac{\pi w}{4}+i\frac{w}{2}\left(\ln{\frac{w}{2}-2\phi_{\rm m}(y)}\right)\right] \\
    &\times\Gamma\left(1-\frac{iw}{2}\right) {}_1 F_{1}\left(i\frac{w}{2},1;\frac{i}{2}w y^2\right) \,,
\eea
where ${}_1 F_1$ is the confluent hypergeometric function,
\be\label{eq:w}
w = 8\pi (1+z_l) M_l f \,,
\ee
and
\be
    \phi_{\rm m}(y) = \frac{(x_{\rm m}-y)^2}{2}-\ln{x_{\rm m}} \,, \qquad
    x_{\rm m} = \frac{y+\sqrt{y^2+4}}{2} \,.
\ee
For a point mass lens the amplification factor $F(f)$ depends only on two parameters: the ``redshifted'' lens mass $M_{lz} \equiv (1+z_l)M_l$ and the rescaled source position $y = |\vec{y}|$. We show the amplification factor~\eqref{pointamplitud} as a function of $w$ for different values of $y$ in Fig.~\ref{fig:amplification}. Two regimes can be seen: In the first one $|F|$ increases monotonically and in the second it oscillates periodically. The two regimes are associated to the wave and geometrical optics regime of the lensing. 

The geometrical optics limit is the limit where the GW wavelength is much smaller than the Schwarzschild radius of the lens. In this regime the amplification factor can be approximated by
\be \label{geometricampli}
    F(f)=| \mu_{+} |^{\frac{1}{2}}-i| \mu_{-}|^{\frac{1}{2}}\exp\left[i2\pi f \Delta t_d\right] \,,
\ee
shown by the dashed colored curves in the right panel of Fig.~\ref{fig:amplification}. In this expression we can see more clearly the effect of the lensing: The lens produces two images of the GW,
\be
    \mu_{\pm} = \frac{1}{2}\pm \frac{2+y^2}{2y\sqrt{y^2+4}} \,,
\ee
with a phase difference proportional to the time delay between the images,
\be\label{timediference}
    \Delta t_d = 4M_{l} \left[\frac{y}{2}\sqrt{y^2+4}+\ln{\left(\frac{\sqrt{y^2+4}+y}{\sqrt{y^2+4}-y}\right)}\right] \,.
\ee
The phase difference between the two images of the source translates into a frequency dependent modulation of the amplitude which gets bigger with the lens mass, with the $y$ parameter and with the GW frequency. This modulation is clearly visible in Fig.~\ref{fig:amplification}. By comparing the solid and dashed curves in Fig.~\ref{fig:amplification} we see that above the first maximum of $|F|$ the geometrical optics gives a good approximation of the amplification factor. Numerically, we find that the first maximum is at
\be
    w \approx \left(0.95 y^{1.8} + 0.37 y^{3.8}\right)^{-0.53} \,.
\ee

\section{Parameter estimation}

We estimate the (unlensed) GW signal from a compact object inspiral by the 2PN expression~(see e.g.~\cite{Cutler:1994ys}):
\be
     \tilde{h}_0(f) = \sqrt{\frac{5}{24}} \frac{\mathcal{M}_z^{5/6}}{\pi^{2/3}D_L} \,f^{-7/6} \,e^{i\Psi(f)}\,,
\ee
where $D_L = (1+z_s^2) D_s$, $\mathcal{M}_z = (1+z_s) \mathcal{M}$ and
\bea
\Psi(f) =& \,2\pi f t_c - \phi_c - \frac{\pi}{4} + \frac{3}{128} (\pi \mathcal{M}_z f)^{-5/3} \\&\times\bigg[1+ \frac{20}{9} \left(\frac{743}{336} + \frac{11}{4}\eta \right) v^2 + (4\beta - 16\pi) v^3 \\& + 10\left(\frac{3058673}{1016064}+\frac{5429}{1008} \eta + \frac{617}{144} \eta^2 - \sigma\right) v^4\bigg] \,,
\eea
with $v = (\pi \mathcal{M}_z \eta^{-3/5} f)^{1/3}$. The unlensed waveform includes in total 7 parameters: the binary chirp mass $\mathcal{M}_z$, symmetric mass ratio $\eta$, and luminosity distance $D_L$, the spin parameters $\beta$ and $\sigma$, the coalescence time $t_c$ and the corresponding phase $\phi_c$. The lensed waveform is simply given by $\tilde h(f) = F(f) \tilde h_0(f)$, and, assuming a point mass lens, it includes 2 additional parameters: the  lens mass $M_{lz}$ and the rescaled source position $y$. For simplicity we don't account for the source sky location, the binary inclination and the signal polarization angles in the analysis, but we instead simply average over them.\footnote{When averaged over the sky location, polarization and inclination angles, a factor $2/5$ needs to be taken into account for LIGO and a factor $3/10$ for ET.} As we consider only the inspiral phase,\footnote{The lensing causes frequency dependent amplification of the signal. Since the merger and ringdown phases cover only a narrow frequency range, the lens parameters can be best measured from the inspiral phase and including the merger and ringdown phases would give only a minor improvement.} we cut the waveform at the frequency corresponding to the radius of the innermost stable circular orbit,
\be \label{eq:fcut}
    f_{\rm cut} = \left[6^{3/2}\pi \mathcal{M}_z \eta^{-3/5} \right]^{-1} \,.
\ee 
The vertical dashed lines in Fig.~\ref{fig:amplification} indicate the value of $w$ corresponding to the cut-off frequency for different values of the lens mass assuming a binary of $20M_\odot$ mass BHs and a luminosity distance $D_L=1.5\rm Gpc$.

\begin{table*}
\caption{Estimated errors of parameter extraction from a GW signal from a BH binary inspiral at $D_l = 1.5 \,$Gpc with component masses $m_1 = m_2 = 20 M_\odot$. The first two rows show the unlensed case and the last two rows the lensed case with $M_{lz} = 100M_\odot$ and $y=1$.}
\label{table1}
\begin{center}
    \begin{tabular}{|c|c|c|c|c|c|c|c|c|c|c|}
    \hline \hline
    & {\rm SNR} & $\frac{\Delta \mathcal{M}_z}{\mathcal{M}_z}[\%]$ & $\frac{\Delta \eta}{\eta}[\%]$ &  $\frac{\Delta D_L}{D_L}[\%]$ & $\Delta \phi$ & $\Delta t_c[s]$ & $\Delta \sigma$ & $\Delta \beta$ & $\frac{\Delta M_{lz}}{M_{lz}}[\%]$ & $\frac{\Delta y}{y}[\%]$ \\ 
    \hline
    LIGO & 46.85 & 0.80 & 144.4 & 2.23 & 3.29 & 0.0003 & 1.61 & 8.26 & - & -  \\
    ET & 388.56 & 0.024 & 9.63 & 0.25 & 0.388 & 0.000043 & 0.16 & 0.48 & - & - \\
    \hline
    LIGO & 58.49 & 0.40 & 1.32 & 3.50 & 2.13 & 0.00025 & 1.59 & 0.77 & 10.39 & 9.74  \\
    ET & 453.86 & 0.0037 & 0.12 & 0.62 & 0.093 & 0.000019 & 0.053 & 0.019 & 1.89 & 1.79 \\
    \hline \hline
    \end{tabular}
\end{center}
\end{table*}

\begin{figure*}
    \vspace{-1.2cm}
    \includegraphics[width=0.48\textwidth]{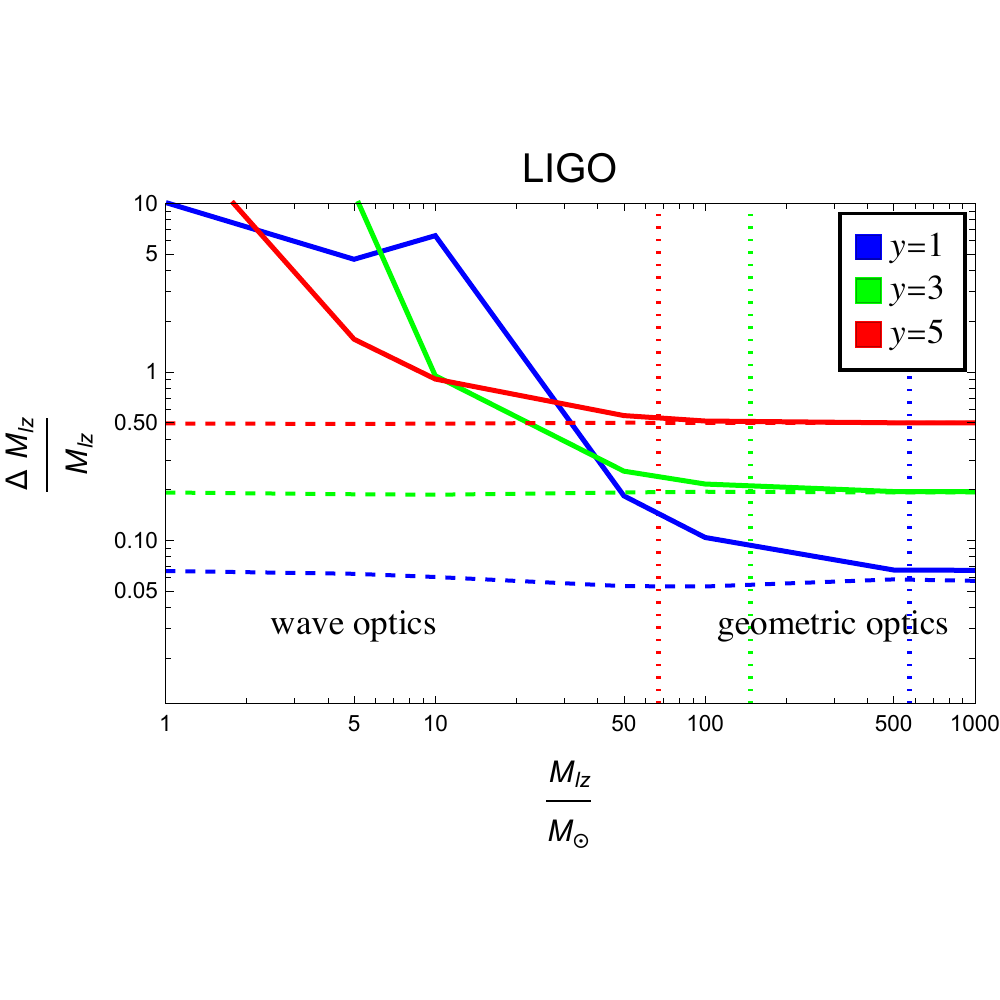} \hspace{2mm}
    \includegraphics[width=0.48\textwidth]{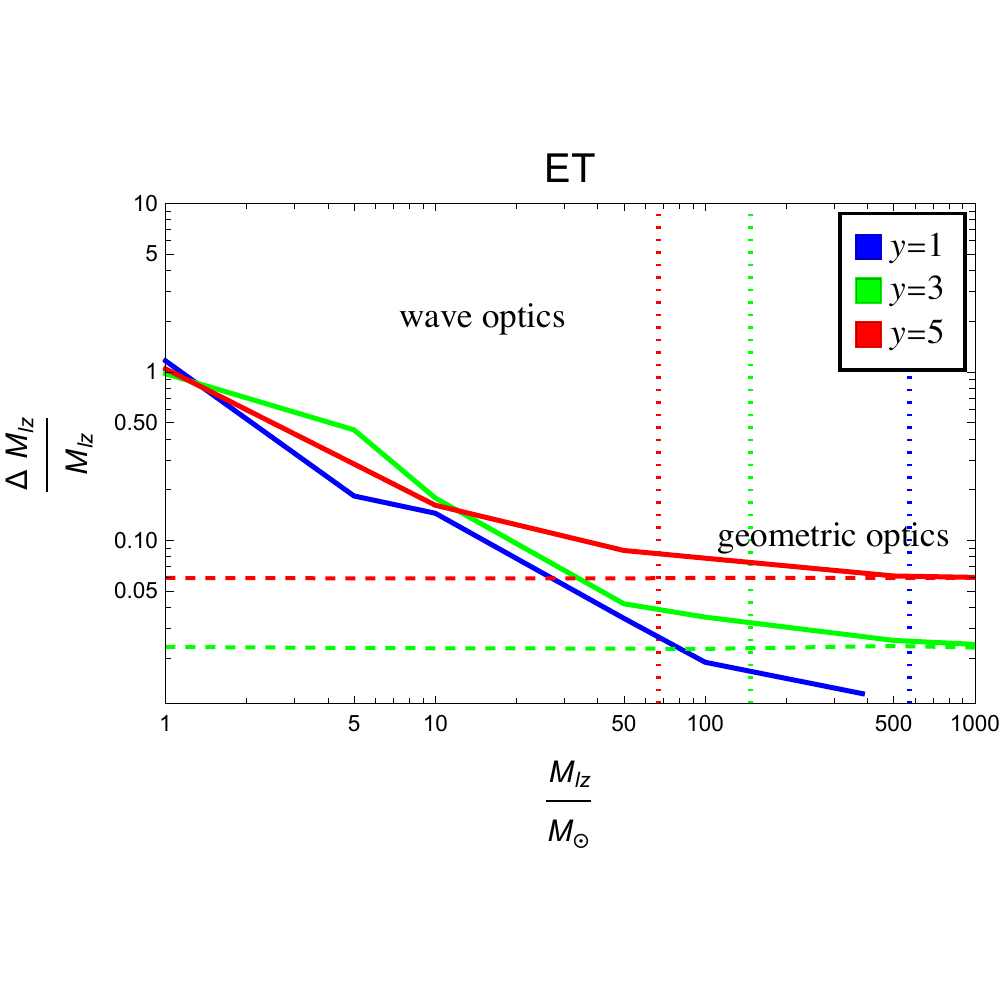} \\ \vspace{-2.2cm}
    \includegraphics[width=0.48\textwidth]{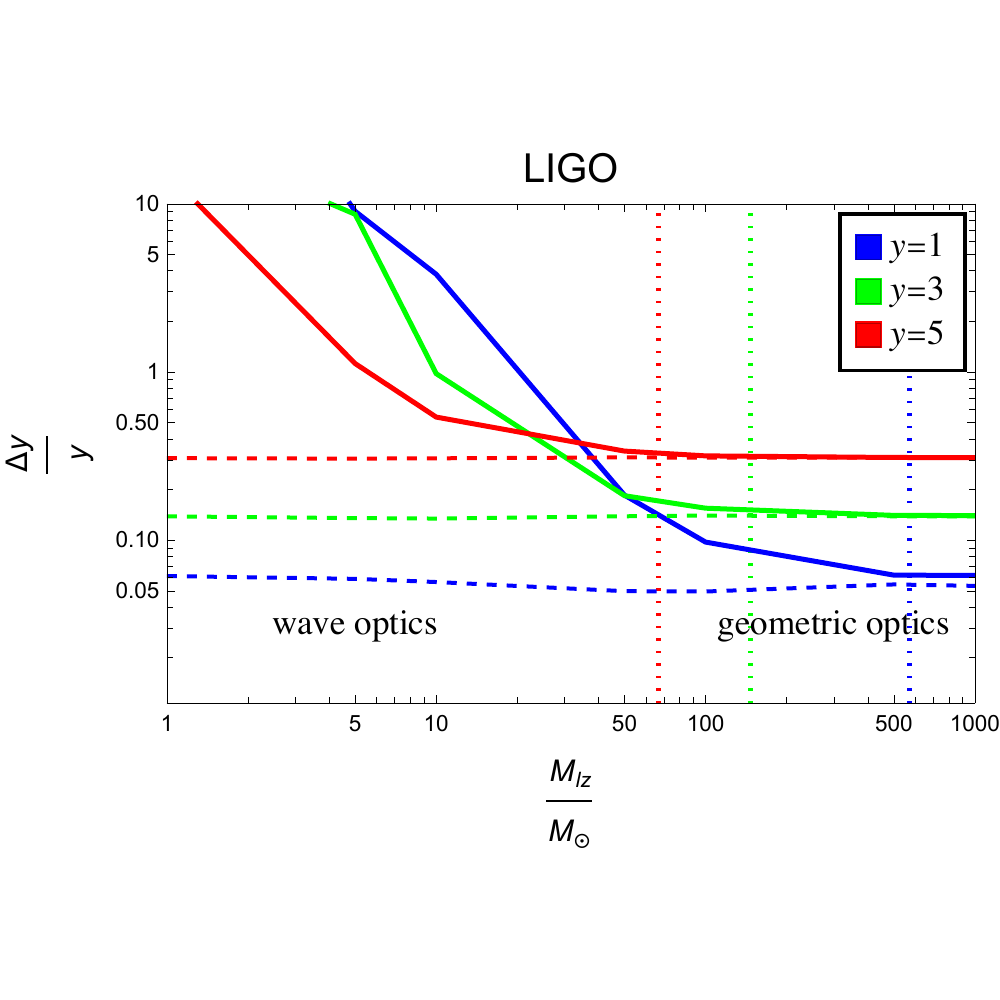} \hspace{2mm}
    \includegraphics[width=0.48\textwidth]{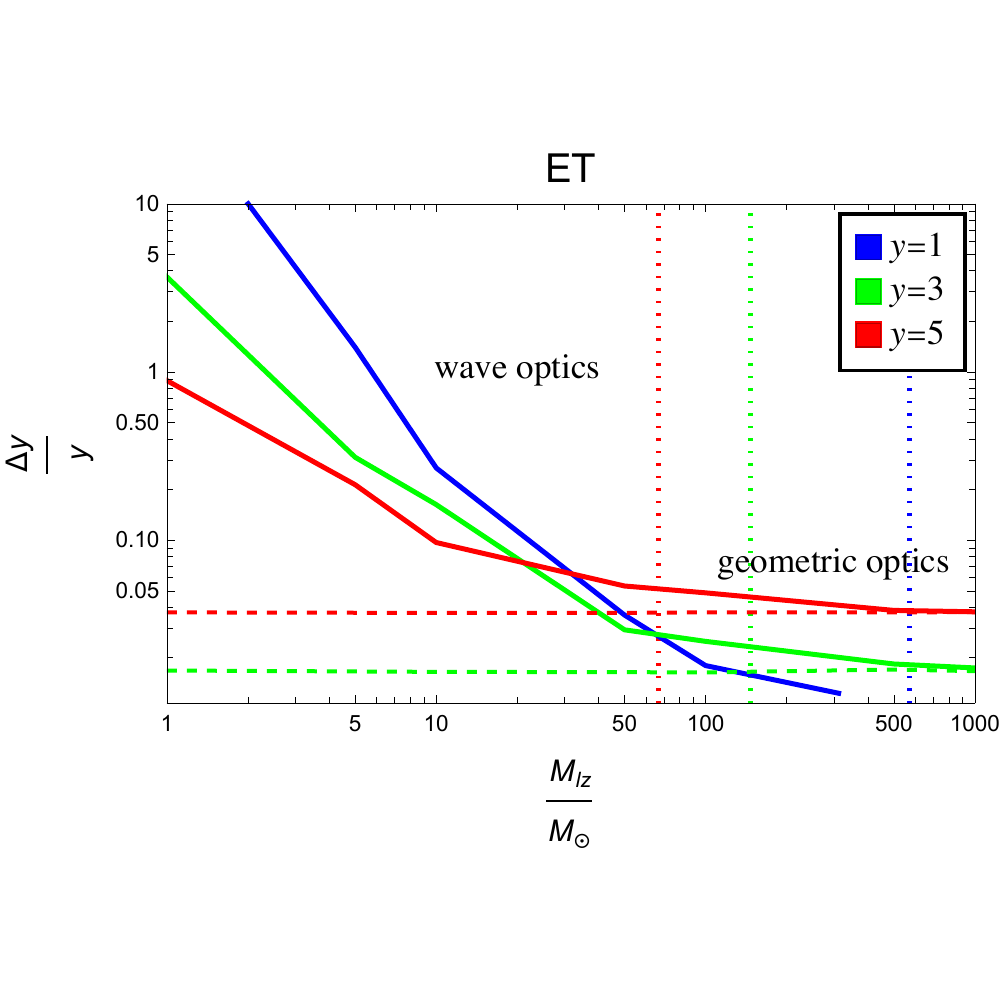} 
    \vspace{-1.2cm}
    \caption{The solid curves show the estimated errors of the lens parameters $M_{lz}$ and $y$ for a signal from a BH binary at $D_l=1.5\,$Gpc ($z_s=0.3$) with component masses $m_1 = m_2 = 20\rm M_{\odot}$. The dashed curves indicate the geometrical optics approximation~\eqref{eq:optics}, and the vertical dotted lines correspond to the estimate~\eqref{eq:Mmin} of the beginning of the validity of the geometrical optics approximation.}
    \label{fig:BH}
\end{figure*}

We use the Fisher analysis (see e.g.~\cite{Poisson:1995ef}) to estimate how accurately the parameters of the source and the lens can be extracted from the waveform.\footnote{Similar analysis has been performed e.g. by \cite{Takahashi:2003ix} and \cite{Cao:2014oaa}.} We start by considering a GW detector whose sensitivity is characterized by the noise power spectral density $S_n(f)$. Defining an inner product for two functions of frequency, $a$ and $b$, as
\be
    \langle a| b \rangle = 2 \int_0^{f_{\rm cut}} \! \td f \,\frac{a^* b + b^* a}{S_n(f)} \,,
\ee
we can express the signal-to-noise ratio as
\be
   {\rm SNR} = \sqrt{\langle \tilde h | \tilde h \rangle} \,,
\ee
which characterizes the detectability of the signal $\tilde h$, and the Fisher matrix $\Gamma_{ij}$ as
\be
    \Gamma_{ij} = \langle \partial_{i}\tilde{h}|\partial_{j}\tilde{h}\rangle \,,
\ee
where the partial derivatives are taken with respect to different parameters of the waveform. For improving the precision of the parameter extraction prior information for the spin parameters can be introduced through the replacement
\be
    \Gamma_{ij} \to  \Gamma_{ij} + \frac{1}{\left(\sigma^{(0)}_{j}\right)^2}\delta_{ij} 
\ee
where following~\cite{Poisson:1995ef}, we have use $\sigma^{(0)}_{\beta}=8.5$ and $\sigma^{(0)}_{\sigma}=5.0$. The inverse of the updated Fisher matrix gives the full covariance matrix, from which we can calculate the standard deviations of the parameter measurements $\Delta j$ and the correlation matrix $c_{ij}$,
\be \label{correlation}
    \Delta j = \sqrt{\Gamma_{jj}^{-1}} \,, \qquad c_{ij} = \frac{\Gamma_{ij}^{-1}}{\Delta i \Delta j} \,.
\ee
For the lens to be detectable we require that $\Delta M_{lz} < M_{lz}/2$ and $\Delta y < y/2$.

Motivated by the LIGO-Virgo observations~\citep{LIGOScientific:2018mvr,LIGOScientific:2020ibl}, we consider a BH binary at angular diameter distance $D_l = 1.5\,$Gpc (corresponding to $z_s \approx 0.3$), with component masses $m_1 = m_2 = 20 M_{\odot}$ and vanishing spin parameters, $\beta = \sigma = 0$. The results of the Fisher analysis for the design sensitivities of LIGO~\citep{LIGOScientific:2014pky} and ET~\citep{Punturo:2010zz} are shown in Table~\ref{table1}. The first two rows correspond to the unlensed case, and the last two rows for a lensed case assuming a lens of mass $M_{lz} = 100M_\odot$ at $y=1$. We see that, due to its amplifying effect, the lensing considerably increases the signal-to-noise ratio of the event and decreases the estimated errors for the source parameters except for the luminosity distance which affects only the amplitude of the signal. For the lens parameters, we see that for this benchmark case the lens is detectable at the design sensitivity of ET and in the design sensitivity of LIGO. 

Our analysis also gives the correlation matrix between the parameters of the waveform both for the lensed and the unlensed case. For example, in the above benchmark case for ET we find the correlation matrix for the lensed case to be
\begin{equation*}
c =
\scalebox{0.63}
{$
\begin{blockarray}{cccccccccc}
 D_l & \mathcal{M}_z & \phi_c & \beta & \sigma & t_c & \eta & M_l & y & \\
 \begin{block}{(ccccccccc)c}
  1.00 & 0.0388 & 0.280 & 0.119 & 0.208 & 0.308 & 0.828 & 0.849 & 0.883 & D_l \\
  0.0388 & 1.00 & 0.844 & 0.922 & 0.900 & 0.687 & 0.0557 & 0.119 & 0.113 & \mathcal{M}_z \\
  0.280 & 0.844 & 1.00 & 0.873 & 0.990 & 0.932 & 0.171 & 0.355 & 0.354 & \phi_c \\
  0.119 & 0.922 & 0.873 & 1.00 & 0.923 & 0.729 & 0.286 & 0.0358 & 0.0117 & \beta \\
  0.208 & 0.900 & 0.990 & 0.923 & 1.00 & 0.888 & 0.0889 & 0.297 & 0.294 & \sigma \\
  0.308 & 0.687 & 0.932 & 0.729 & 0.888 & 1.00 & 0.242 & 0.336 & 0.352 & t_c \\
  0.828 & 0.0557 & 0.171 & 0.286 & 0.0889 & 0.242 & 1.00 & 0.617 & 0.678 & \eta \\
  0.849 & 0.119 & 0.355 & 0.0358 & 0.297 & 0.336 & 0.617 & 1.00 & 0.984 & M_l \\
  0.883 & 0.113 & 0.354 & 0.0117 & 0.294 & 0.352 & 0.678 & 0.984 & 1.00 & y \\
 \end{block}
\end{blockarray}
$} \,.
\end{equation*}
As expected, from the source parameters, the luminosity distance has the strongest correlation with the lens parameters. This is because the biggest effect of the lensing is on the amplitude of the waveform which is controlled by the luminosity distance.

In Fig.~\ref{fig:BH} we show the estimated relative errors of $M_{lz}$ and $y$ as a function of $M_{lz}$ for different values of $y$ for the same source parameters as considered in Table~\ref{table1}. We can see that in this benchmark case LIGO can detect a lens mass $M_{lz} \gtrsim 30 \rm M_{\odot}$ whereas ET can detect $M_{lz} \gtrsim 1 \rm M_{\odot}$. The relative errors fluctuate and they do not decrease monotonically with the lens mass near the transition from the wave optics regime to the geometrical optics regime. This is a consequence of diffraction which makes the amplitude to fluctuate with the frequency, as shown in Fig.~\ref{fig:amplification}.

In the geometrical optics limit approximate analytical expressions for the estimated measurement errors of the lens parameters were derived in~\citep{Takahashi:2003ix}:
\bea \label{eq:optics}
    &\frac{\Delta M_{lz}}{M_{lz}} = \frac{1}{\rm SNR} \frac{\sqrt{y(y^2+2)}(y^2+4)^{\frac{5}{4}}}{2\tau} \,, \\
    &\frac{\Delta y}{y} = \frac{1}{\rm SNR} \frac{\sqrt{y^2+2}(y^2+4)^{\frac{3}{4}}}{2\sqrt{y}} \,,
\eea
where $\tau = \Delta t_d/(4M_{lz})$. In particular, these expressions show that the lens parameters can be determined with an accuracy inversely proportional to the signal-to-noise ratio. These approximations are shown in Fig.~\ref{fig:BH} by the dashed curves. We see that for high lens masses, the estimated errors in $M_{lz}$ and $y$ converge to constant values given by the geometrical optics approximations. To estimate the lower bound on lens mass above which the geometrical optics approximations hold, we require that $|F|$ oscillates at least for 2.5 periods before the cut-off frequency~\eqref{eq:fcut}. Numerically we find that the 5th root of $|F|-1$ is at 
\be
    w \approx \left(0.018 y^{2.0} + 0.0014 y^{3.9}\right)^{-0.52} \,.
\ee
Requiring that this corresponds to a frequency lower than $f_{\rm cut}$ we get a lower bound on the lens mass, 
\be \label{eq:Mmin}
    M_{lz} \gtrsim \frac{1.84 \mathcal{M}_z \eta^{3/5} }{\left(0.018 y^{2.0} + 0.0014 y^{3.9}\right)^{0.52}} \equiv M_{\rm min} \,.
\ee
As indicated by the vertical dotted lines in Fig.~\ref{fig:BH}, this gives a good approximation of the minimal lens mass for which the geometrical optics approximations hold. We note that this lower bound decreases for increasing values of $y$, and is proportional to the total mass $M_z = \mathcal{M}_z \eta^{3/5}$ of the GW source.

\section{Lensing probability}

Next we will derive the optical depth for a GW source at a given distance and the expected number of detectable lens events. We consider a population of compact lens objects distributed uniformly with number density 
\be
    n_l = (1+z_l)^3 \frac{f_{\rm DM} \rho_{\rm DM}}{M_l}\,,
\ee
where $\rho_{\rm DM} \approx 33 M_{\odot}/{\rm kpc}^3$~\citep{Planck:2018vyg} denotes the present average DM density and $f_{\rm DM}$ the fraction of DM in the lens objects. For a lens at redshift $z_l$ the lensing cross section is 
\be
    \sigma = \pi y_c^2 \theta_{E}^2 D_l^2 \,,
\ee
where $\theta_E$ is the Einstein ring radius given in Eq.~\eqref{Einsteinradius}, and $y_c$ is the maximal value of the $y$ parameter for which the lens can be detected ($\Delta y/y < 1/2$ for $y>y_c$). Notice that the cross section is enhanced by the factor $y_c^2$ compared to lensing of electromagnetic signals under the strong lensing condition, where $y_c$ is normally assumed to be 1.

In the geometrical optics limit, neglecting the subdominant dependence of the signal-to-noise ratio on the lens properties, $y_c$ depends only on the properties of the source, as can be seen from Eq.~\eqref{eq:optics}. The lensing optical depth for a signal from a source at redshift $z_s$ is then obtained by integrating over the lens redshift,
\bea \label{eq:depth}
    \tau_l &= \int_0^{z_s} \td z_l\, \frac{\sigma(z_l) n_l(z_l)}{(1+z_l)H(z_l)} \\
    &= 4\pi f_{\rm DM} \rho_{\rm DM} y_c^2 \int_{0}^{z_s} \td z_l \,\frac{(1+z_l)^2}{H(z_l)} \frac{D_l D_{ls}}{D_s} \,,
\eea
where $H$ denotes the Hubble expansion rate. Assuming Poisson distribution for the lens events, the probability that the signal is lensed (at least once) is given by
\be \label{eq:Pl}
    P_l = 1 - e^{-\tau_l} \,. 
\ee 
By numerically evaluating~\eqref{eq:depth}, we find that 
\be
P_l\approx \min[0.07 f_{\rm DM} y_c^2 z_s^2,\,1] \,.
\ee
In particular, for $z_s^2 \gg 14/(f_{\rm DM} y_c^2)$ lensing is almost certain. For the following results we will evaluate Eq.~\eqref{eq:Pl} numerically. 

The LIGO-Virgo detectors have during the observation runs O1, O2 and O3a observed in total 50 compact object coalescence events~\citep{LIGOScientific:2018mvr,LIGOScientific:2020ibl}. Assuming that some fraction of DM consists of compact objects, we would expect to see 
\be
N_l = \sum \theta(M_{lz} - M_{\rm min}) P_l \,.
\ee
lensed events. Here the sum runs over the 50 events, the Heaviside theta function accounts for the minimal observable lens mass which we estimate using Eq.~\eqref{eq:Mmin}, and $M_{\rm min}$ and $P_l$ are evaluated for the central values of the measured masses and distances of the binaries. Moreover, knowing the merger rate $R$ of compact object binaries (see appendix~\ref{app:rate}) we can estimate the expected number of detectable lens events for a future observatory within the observation time $\mathcal{T}$ as
\be \label{eq:Nlens}
    n_l = \mathcal{T} \int \td \lambda\, \theta(M_{lz} - M_{\rm min}) P_l\, p_{\rm det}({\rm SNR}_c/{\rm SNR}) \,,
\ee
where $p_{\rm det}$ is the detection probability averaging over the binary sky location and inclination, and the polarization of the signal~\citep{Gerosa:2019dbe}, and
\be
    \td \lambda = \frac{1}{1+z_s}\frac{\td R}{\td m_1 \td m_2} \frac{\td V_c}{\td z_s} \td m_1 \td m_2 \td z_s \,.
\ee
For the threshold signal-to-noise ratio we take ${\rm SNR}_c = 8$. The expected total number of detectable events is instead given by
\be
    N_{\rm tot} =  \mathcal{T} \int \td \lambda \, p_{\rm det}({\rm SNR}_c/{\rm SNR}) \,.
\ee

\section{Results}

\begin{figure}
  \centering
  \includegraphics[width=0.46\textwidth]{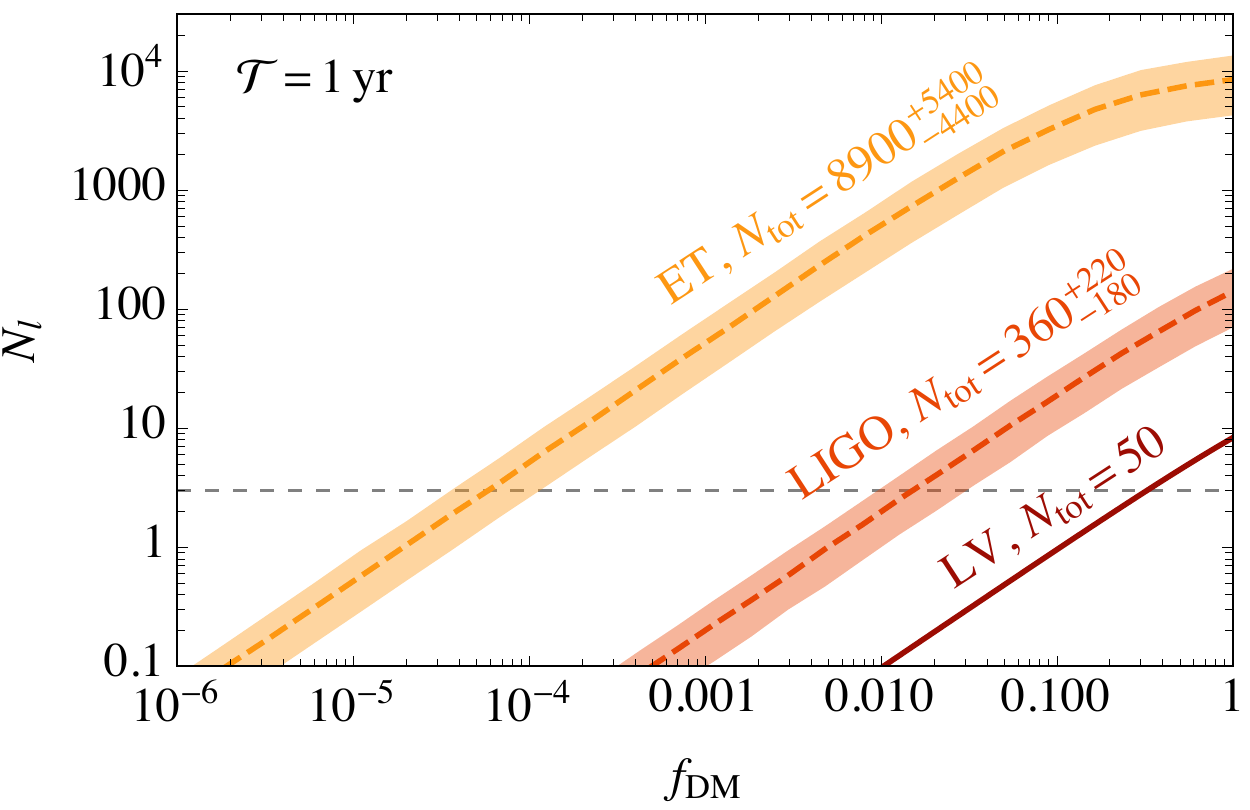}
  \caption{Expected number of detectable lensed GW events as a function of the fraction of DM in the lens objects, assuming that the lens objects are sufficiently heavy such that the lensing can be described by geometrical optics. The width of the bands indicate the uncertainties in the merger rate, and the expected total numbers of detectable GW events are shown in the plot.}
  \label{fig:Nlens}
\end{figure}

Let us first consider a population of heavy lenses for which the geometrical optics approximation holds for the whole detectable source population. In Fig.~\ref{fig:Nlens} we show by the solid red curve the expected number of lensed GW events considering the 50 compact object binary coalescences seen during LIGO-Virgo observation runs O1, O2 and O3a~\citep{LIGOScientific:2018mvr,LIGOScientific:2020ibl}, and by the dashed orange and yellow curves the projections of the LIGO design sensitivity and ET assuming an observations time of one year, $\mathcal{T} = 1\,$yr. We see that, remarkably, even in its current sensitivity, the LIGO-Virgo detectors should have seen 8 lensed events if all DM would be in such heavy objects. The search of lensed events in the LIGO-Virgo data, however, indicates that none of the events so far has been lensed~\citep{LIGOScientific:2021izm}.\footnote{We note that evidence for lensed events in the LIGO-Virgo data was recently claimed by~\cite{Diego:2021fyd}.} We therefore place a $95\%$ upper bound on $f_{\rm DM}$ by requiring that $N_l < 3$.\footnote{The LIGO-Virgo analysis is based on evidence ratio between lensed and unlensed hypotheses, while our work uses parameter estimation and we assume that the lensed waveform can be separated from the unlensed one if the lens parameters are more than $2\sigma$ away from $0$. We leave for future studies the analysis of how well these approaches match.} This implies that $f_{\rm DM} < 0.4$. Moreover, we see that in its design sensitivity LIGO can probe the abundance of the lens objects down to $f_{\rm DM} = 0.02$ and ET down to $f_{\rm DM} = 7\times 10^{-5}$. If the heavy compact objects constitute more than $10\%$ of DM then almost all GW events that ET would see would be lensed. The width of the LIGO and ET bands in Fig.~\ref{fig:Nlens} reflect the uncertainties in the merger rate. These uncertainties will reduce in near future as the merger rate will be more accurately measured.

\begin{figure}
  \centering
  \includegraphics[width=0.48\textwidth]{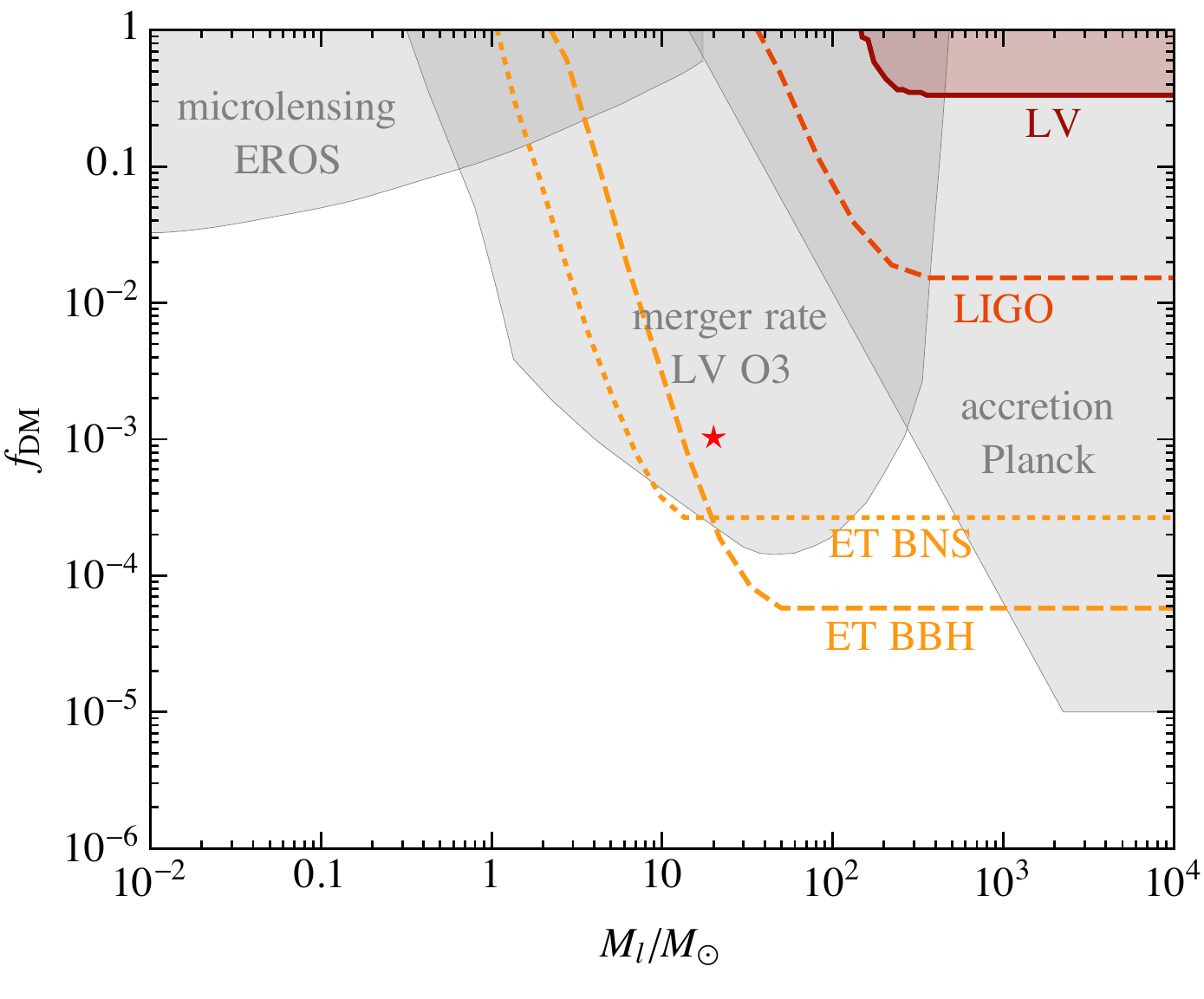}
  \vspace{-5mm}
  \caption{The solid red curve shows the current LIGO-Virgo constraint on the abundance of compact objects arising from the non-observation of lensed GW events. The dashed orange and yellow curves instead indicate the projected sensitivities of LIGO in its design sensitivity and ET for probing compact objects trough lensing of the GW events. The gray shaded regions show the current constraints for compact DM objects arising from the EROS microlensing observations~\citep{EROS-2:2006ryy}, and the constraints on PBHs arising from the BH binary merger rate observed by LIGO-Virgo~\citep{Hutsi:2020sol}, and from the Planck CMB constraints on accreting PBHs~\citep{Ali-Haimoud:2016mbv,Serpico:2020ehh}. The red star indicates the scenario where all LIGO-Virgo BHs are primordial~\citep{Hutsi:2020sol}.}
  \label{fig:constr}
\end{figure}

In Fig.~\ref{fig:constr} we extend the results for different lens masses\footnote{We consider monochromatic mass distribution of the compact DM objects. Neglecting possible clustering effects, our results can be extended to broader mass functions using the method derived by~\cite{Carr:2017jsz}.} The solid red curve indicates the current LIGO-Virgo $N_l < 3$ constraint on the abundance of the lens objects, $f_{\rm DM}$, as a function of their mass $M_l$. Other constraints in the shown mass range arise from microlensing of stellar light~\citep{Griest:2013aaa,Macho:2000nvd,EROS-2:2006ryy,Niikura:2017zjd,Niikura:2019kqi}, lensing of type Ia supernovae~\citep{Zumalacarregui:2017qqd}, the BH merger rate observed by LIGO-Virgo~\citep{Raidal:2017mfl,Ali-Haimoud:2017rtz,Raidal:2018bbj,Vaskonen:2019jpv,DeLuca:2020qqa,Hutsi:2020sol}, PBH accretion~\citep{Ricotti:2007au,Ali-Haimoud:2016mbv,Poulin:2017bwe,Hektor:2018qqw,Serpico:2020ehh}, survival of a stars in dwarf galaxies~\citep{Brandt:2016aco,Koushiappas:2017chw}, Lyman-$\alpha$ forest data~\citep{Afshordi:2003zb,Murgia:2019duy}, and strong lensing of fast radio bursts~\citep{Munoz:2016tmg,Liao:2020wae}. For clarity, only the strongest of these constraints are shown in Fig.~\ref{fig:constr}. We note that the constraints other than the lensing constraints assume that the compact DM objects are BHs.

We see that the LIGO-Virgo lensing constraint excludes DM fractions larger than $0.4$ for masses $M_l \gtrsim 400M_\odot$, and the scenario that massive compact objects constitute all DM for $M_l > 200M_\odot$. We also show the expected reaches of the LIGO at its design sensitivity and ET with dashed orange and yellow curves using the central value for the present merger rate, $R_0 = 10\,{\rm Gpc}^{-3} {\rm yr}^{-1}$. These experiments will, respectively, probe the scenario where massive compact objects constitute all DM for $M_l > 40M_\odot$ and $M_l > 2M_\odot$, and DM fractions of $0.02$ and $7\times 10^{-5}$. In particular, ET will be able to probe the scenario where all LIGO-Virgo BHs are primordial (red star in Fig.~\ref{fig:constr}). Moreover, we find that that by searching for lensed GW signals from NS binaries only ET can probe the abundance of the lens objects. This prospect is indicated in Fig.~\ref{fig:constr} by the yellow dotted curve. We see that, while the reach is not as good at high masses as in the case of the lensing of BH binary signals, slightly lighter lens objects, $M_l > 1 M_\odot$, can be probed. This is simply because the merger frequency is higher for lighter binaries, and therefore the minimal detectable lens mass is lower for the NS binary signals than for BH binary signals.

\section{Conclusions}

We have shown that compact DM objects can be probed through lensing of GW signals. We have used the Fisher analysis to estimate the accuracy at which different parameters of the lensed waveform can be measured, finding that, for example, from a GW signal that originates from a BH binary with $m_1 = m_2 = 20M_\odot$ at luminosity distance $D_L = 1.5\,{\rm Gpc}$, the effect of the lens can be detected with LIGO in its design sensitivity if the lens is heavier than $100M_\odot$ and with ET if it is heavier than $10M_\odot$, assuming that the source and the lens are sufficiently closely aligned on the sky. The lensing also increases the signal-to-noise ratio of the event and decreases the estimated errors of the source parameters except for its luminosity distance. Moreover, we have shown that the maximal misalignment between the source and the lens (that is, the maximal detectable value of $y$) can be well approximated by geometrical optics, and the minimal detectable lens mass can be estimated by the minimal lens mass for which the geometrical optics description works.

Using the results of the Fisher analysis, we have then estimated the optical depth for a signal from a given source assuming a homogeneous distribution of compact lens objects, and the probability that a given GW event is lensed by a detectable amount. Considering as the GW sources the $\mathcal{O}(10M_\odot)$ BH binary population whose merger rate we know from the LIGO-Virgo observations, we have derived an upper bound on the abundance of the compact lens objects from non-detection of lensed GW events by LIGO-Virgo. This bound excludes the scenario that all DM consists of compact objects heavier than $200M_\odot$ and reaches DM fraction of $0.4$ for masses heavier than $400M_\odot$. We have also estimated the prospects for the LIGO design sensitivity and ET. We have shown that these detectors will greatly extend the mass and abundance range of compact DM objects that can be probed through lensing of GW signals from BH binaries. In particular, ET will reach down to $M_l \sim 1M_\odot$ and $f_{\rm DM} \sim 10^{-4}$.

Our analysis does not account for structures formed by the lens objects, and neglects macrolensing effects~\citep{Diego:2019rzc} and possible astrophysical lenses~\citep{Xu:2021bfn}. We leave the studies of these effects for future work. Another interesting future direction would be to consider lensing by objects such as axion miniclusters~\citep{Fairbairn:2017sil} or dark low-mass halos~\citep{Oguri:2020ldf} which can not be approximated as point masses.

\section*{Acknowledgments} 
We thank Hardi Veerm\"ae for valuable comments. This work was supported by the Spanish MINECO grants FPA2017-88915-P and SEV-2016-0588, the grant 2017-SGR-1069 from the Generalitat de Catalunya. IFAE is partially funded by the CERCA program of the Generalitat de Catalunya.

%\section*{Data availability}
%The data underlying this article will be shared on reasonable request to the authors.

\bibliographystyle{mnras}
\bibliography{references}

\appendix

\section{Merger rate}
\label{app:rate}

As the GW sources in this work we consider the BH and NS binary populations probed by the LIGO-Virgo observations~\citep{LIGOScientific:2020kqk}. In this appendix we review their merger rate. For the purpose of this work sophisticated modeling of the source population is not needed. Instead, we adapt the simple model used e.g. by~\cite{Hutsi:2020sol}: Following \cite{Belczynski:2016obo}, we assume that the merger rate inherits the redshift dependence of the star formation rate~\citep{Madau:2016jbv},
\be
	{\rm SFR}(z) \propto \frac{(1 + z)^{2.6}}{1 + ((1 + z)/3.2)^{6.2}} \equiv P_b(z)\,,
\ee 
with a delay $P_d(t) \propto t^{-1}$ with $t > 50\,{\rm Myr}$. At $z\lesssim 2$ this model predicts merger rate $R\propto (1+z)^{1.6}$, which is in an excellent agreement with observations~\citep{LIGOScientific:2020kqk}. We note, however, that the experimental uncertainties in the scaling of the merger rate with redshift are large, and this may impact the prospects of probing the lens population with with the future more sensitive experiments~\citep{Mukherjee:2021qam,Wang:2021lij}. The differential merger rate is
\bea \label{eq:RABH}
	&\frac{\td R}{\td m_1 \td m_2} = \frac{R_0}{Z_\psi} M^\alpha \eta^\beta \psi(m_1)\psi(m_2) \\
	&\times \int \td t_d \td z P_b(z) P_d(t_d) \delta(t(z_s)-t(z)-t_d) \,,
\eea
where $\alpha$ and $\beta$ parametrize the mass dependence of the binary population, the normalization factor $Z_\psi$ is defined such that $R(z_s=0) = R_0$. The mass function for the compact object population is denoted by $\psi$. For BHs we take a truncated power-law mass function,
\be
    \psi(m) \propto m^{\zeta} \, \theta(m - m_{\rm min})\theta(m_{\rm max}- m) \,,
\ee
with the normalization $\int \psi(m) \td\ln m = 1$. The choice upper an lower mass limits, $m_{\rm min} = 3.0 M_\odot$ and $m_{\rm max} = 55 M_\odot$, arise from the maximal NS mass and the pair-instability supernovae. The present BH merger rate $R_0$ and the powers $\alpha$, $\beta$, and $\zeta$ are fixed by the LIGO-Virgo observations~\citep{LIGOScientific:2018mvr,LIGOScientific:2020ibl}, which indicate $\alpha = 0$ and $\beta = 6$, $\zeta = -0.5$, and $R_0 = 10^{+6}_{-5}\,{\rm Gpc}^{-3} {\rm yr}^{-1}$~\citep{Hutsi:2020sol}. For NS binaries we instead take a monochromatic mass function at $1.4M_\odot$, and $R_0 = 320\,{\rm Gpc}^{-3}{\rm yr}^{-1}$~\citep{LIGOScientific:2020kqk}.

\label{lastpage}

\end{document}